\documentclass[twocolumn,           
               showpacs,            
               preprintnumbers,     
               aps,                 
               prd,                 
               a4paper,             
               superscriptaddress,  
               nofootinbib,         
               tightenlines,        
               floats,floatfix      
               ]{revtex4}
\usepackage{graphicx}
\usepackage{latexsym}
\usepackage{amsmath,amssymb}        
\usepackage[draft=false]{hyperref}

\begin{document}



\title{Constraining slow-roll inflation with WMAP and 2dF}
\author{Samuel M.~Leach}
\affiliation{D\'epartement de Physique Th\'eorique, Universit\'e de Gen\`eve,
24 quai Ernest Ansermet, CH-1211 Gen\`eve 4, Switzerland}
\author{Andrew R.~Liddle}
\affiliation{Astronomy Centre, University of Sussex, Brighton BN1 9QJ, United
Kingdom}
\date{\today}
\pacs{98.80.Cq \hfill astro-ph/0306305}
\preprint{astro-ph/0306305}


\begin{abstract}
We constrain slow-roll inflationary models using the recent WMAP
data combined with data from the VSA, CBI, ACBAR and 2dF
experiments. We find the slow-roll parameters to be $0<\epsilon_1<
0.032$ and $\epsilon_2+5.0\epsilon_1 = 0.036 \pm 0.025$. For inflation
models $V\propto\phi^{\alpha}$ we find that $\alpha< 3.9,\;4.3$ at
the 2$\sigma$ and $3\sigma$ levels, indicating that the
$\lambda\phi^4$ model is under very strong pressure from
observations. We define a convergence criterion to judge the
necessity of introducing further power spectrum parameters such as
the spectral index and running of the spectral index. This
criterion is typically violated by models with large negative
running that fit the data, indicating that the running cannot be
reliably measured with present data.
\end{abstract}

\maketitle


\section{Introduction}

The observations by the WMAP satellite
\cite{wmap1,wmapS,wmapinf,wmap2,wmap3} have brought the global
cosmological data set up to a quality where, for the first time,
it is possible to obtain precision constraints on cosmological
models. That the data provides no indication of any significant
departure from gaussianity, adiabaticity, or scale-invariance, and
furthermore reveals two coherent peaks in the spectrum of cosmic
microwave background (CMB) anisotropies, lends powerful support to
the idea of inflation in the early Universe as a source for the
observed perturbations. This opens the prospect of constraining
and excluding regions of inflation model parameter space.

The qualitative breakthrough of the WMAP data is important. For
the first time we have available a CMB spectrum that spans from
cosmic variance limited measurements on large angular scales
across to the measurement without any overall calibration error of
the peaks on small angular scales. Measurements of the
polarization of the CMB~\cite{wmap4} provide some insight into the
epoch of reionization which in turn helps to constrain
inflationary models by limiting the effects of parameter
degeneracies.

In this paper we analyze slow-roll inflation models, following the
strategy outlined in Leach et al.~\cite{LLMS}, and provisionally
applied to pre-WMAP CMB data by Leach and Liddle \cite{infcmb} as
a demonstration of methodology. With WMAP it is possible to make
the first serious application. As compared to our earlier work, we
make several improvements. We use the now-ubiquitous Markov Chain
Monte Carlo method \cite{MCMC,LB} to obtain the likelihood function over
parameter
space, we include both short-scale CMB data and the galaxy power
spectrum data from 2dF (but not any lyman-alpha data, whose
inclusion has proven controversial \cite{Sel}), and we study the
effect of varying one further slow-roll parameter. Our approach
differs in several respects from the other papers that have
already appeared discussing inflation post-WMAP
\cite{wmapinf,BLM,KKMR}, and we contrast our work with theirs in
the conclusions.

\section{Methodology}

We follow the methodology described in Refs.~\cite{LLMS,infcmb},
to which the reader is referred for further details. In terms of
the horizon-flow parameters $\epsilon_1$, $\epsilon_2$, etc., the
inflationary scalar and tensor power spectra can be well
represented as power-laws with amplitude and spectral indices
given by~\cite{SL,SL2}
\begin{eqnarray}
A_{{\rm S}} &=& \frac{H^2}{\pi\epsilon_1 m_{{\rm
PL}}^2}\left(1-2(C+1)\epsilon_1-C\epsilon_2 \right)\label{eq:amp}\\
n_{{\rm S}} -1 &=& -2\epsilon_1-\epsilon_2\label{eq:ns}\\
n_{{\rm T}} &=& -2\epsilon_1 \,,
\end{eqnarray}
where $C \simeq -0.73$. The
relative amplitude of tensor and scalar perturbations is given by
\begin{equation}
R = 16\epsilon_1 \,.\label{eq:R}
\end{equation}
Later we will also consider weak running (scale-dependence) of the
scalar spectral index. Although the full inflationary predictions
are somewhat more detailed, Eqs.~(\ref{eq:amp})--(\ref{eq:R})
capture the essence of the inflationary power spectra. In the
following analysis we start by using the first-order power-law
shape predictions for inflation which include the Stewart--Lyth
correction to the amplitude, but ignore any term
$\mathcal{O}(\epsilon^2)$. Later we use the full second-order
predictions where the terms $\mathcal{O}(\epsilon^2)$ are included
in the fit.

The data that we use in this paper comes from VSA \cite{vsa}, CBI
\cite{cbi}, ACBAR \cite{acbar}, WMAP \cite{wmapinf} and the 2dF
galaxy redshift survey \cite{2df}. We compute the microwave
anisotropies using the CAMB code \cite{LCL} coupled to our own
slow-roll inflation module \cite{LLMS}, and use the package
CosmoMC \cite{LB}, modified to include the WMAP likelihood code
\cite{V}, in order to compute the likelihood over parameter space.
We generate a Markov chain of 60,000 elements. We assume a flat
$\Lambda$CDM universe and adopt the parameter basis
$\left\{\omega_{{\rm B}},\omega_{{\rm D}},H_0,10^{10}A_{{\rm
S}}(k_*),z_{{\rm re}},\epsilon_1(k_*),\epsilon_2(k_*)\right\}$
where ${\omega_{{\rm B}}}$ and ${\omega_{{\rm D}}}$ are the baryon
and dark matter physical densities, $H_0$ is the Hubble constant,
$A_{{\rm S}}$ is the amplitude of scalar perturbations, $z_{{\rm
re}}$ is the redshift of reionization (which is assumed to be
instantaneous), and $k_*=0.01{\rm Mpc}^{-1}$. The only prior that
has any effect on the constraints is insisting $z_{{\rm re}}>4$.
We are also interested in the constraints on the derived
parameters $\{A_{{\rm S}}e^{-2\tau},n_{{\rm S}}-1,R_{10} \}$ where
$\tau$ is the optical depth to the last scattering surface and
$R_{10}=C^{{\rm T}}_{10}/C^{{\rm S}}_{10}$.

\begin{figure}[t]
\includegraphics[width=\linewidth]{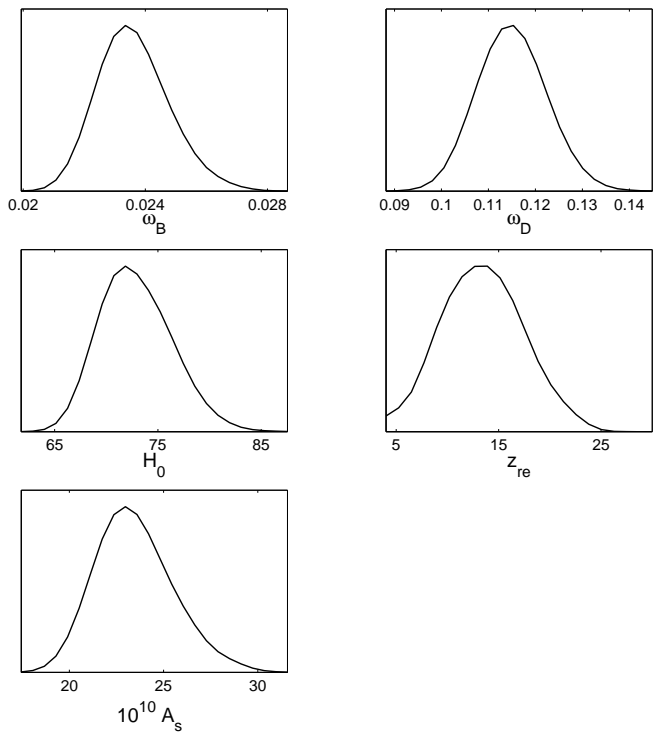}\\
\caption[fig2]{\label{fig:cos_params1} 1D posterior constraints for
the basic cosmological parameters assuming slow-roll inflation.}
\end{figure}

\section{The constraints}

\subsection{Constraints on slow-roll inflation}

We begin by considering the constraints on slow-roll inflation
models, initially only including the horizon-flow parameters
$\epsilon_1$ and $\epsilon_2$. For orientation and comparison with
other work, in Fig.~\ref{fig:cos_params1} we display the
constraints on the basic cosmological parameters
$\left\{\omega_{{\rm B}},\omega_{{\rm D}},H_0,10^{10}A_{{\rm
S}}(k_*),z_{{\rm re}}\right\}$. Our results are in good agreement
with other authors, unsurprisingly as we use many of the same
codes and a similar data compilation.

\begin{figure}[t]
\includegraphics[width=\linewidth]{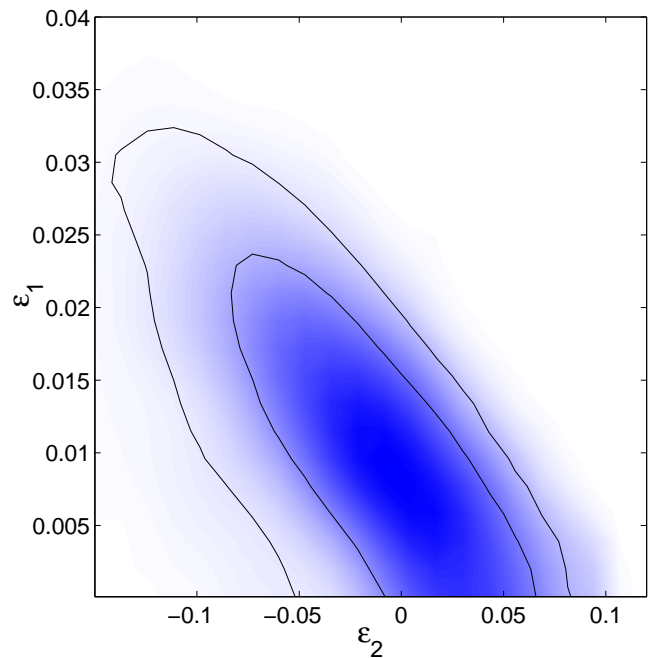}\\
\caption[fig0]{\label{fig:e1e2} 2D posterior constraints in the
$\epsilon_1$--$\epsilon_2$ plane. The contours are the $1\sigma$ and
$2\sigma$ bounds.}
\end{figure}

\begin{figure}[t]
\includegraphics[width=\linewidth]{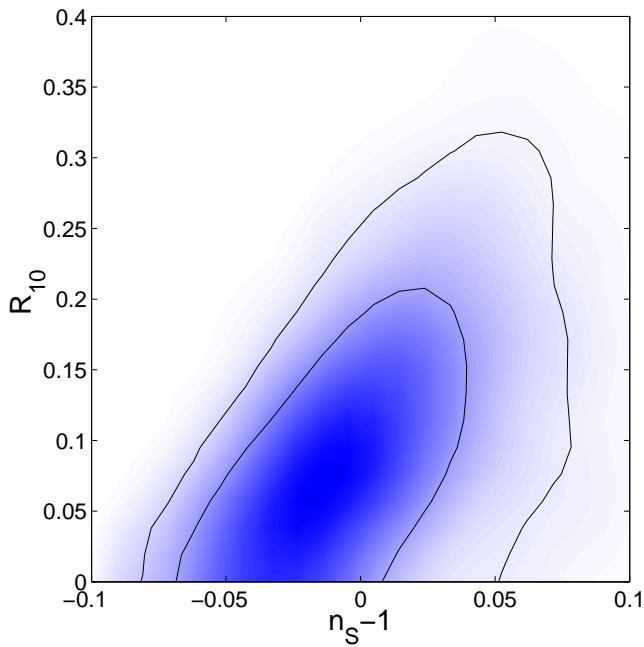}\\
\caption[fig1]{\label{fig:nsR} 2D posterior constraints in the
$(n_{{\rm s}}-1)$--$R_{10}$ plane, again at $1\sigma$ and $2\sigma$. Models with 
a tensor spectrum on large
scales require a bluer scalar spectrum in order to increase CMB
power to short scales.}
\end{figure}

Fig.~\ref{fig:e1e2} shows the likelihood distribution in the plane
of the horizon-flow parameters $\epsilon_1$ and $\epsilon_2$. We
can see that the constraint on $\epsilon_2$ is highly correlated
with $\epsilon_1$, since both parameters contribute to the
spectral index. Moreover the data introduce a further degeneracy,
the tensor degeneracy. This occurs where models with a tensor
component, and hence more power on large scales, require more
power to short scales, and hence a bluer spectrum. This is clear
from Fig.~\ref{fig:nsR}, in which we plot the same constraints in
terms of the derived parameters $n_{{\rm S}}$ and $R_{10}$.

\begin{figure}[t]
\includegraphics[width=\linewidth]{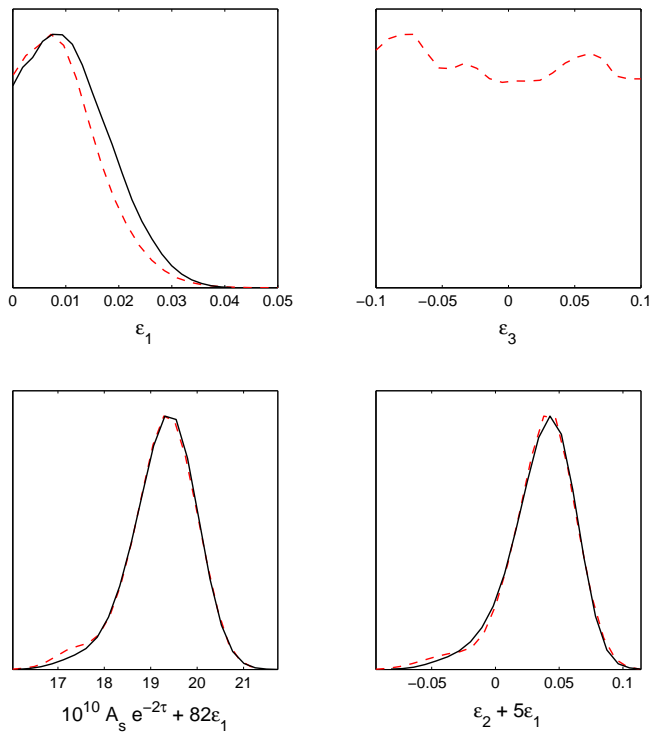}
 \caption[fig1]{\label{fig:e3} 1D posterior constraints on inflationary
parameters. The solid line corresponds to a power-law fit to the
data using $\epsilon_1$ and $\epsilon_2$. The dashed line
corresponds to a fit where weak running of the spectral index is
included in the fit via the slow-roll parameter $\epsilon_3$,
which is unconstrained by the data.}
\end{figure}

Because of the strong degeneracy between $\epsilon_1$ and
$\epsilon_2$, we read off the 2$\sigma$ upper limit on
$\epsilon_1$ from Fig.~\ref{fig:e1e2} without marginalizing out
$\epsilon_2$, finding
\begin{equation}
\epsilon_1< 0.032,\label{eq:eps1const}
\end{equation}
which gives the best measure of the relative (primordial) contribution
of tensors, via Eq.~(\ref{eq:R}). This constraint is in agreement
with Refs.~\cite{wmapinf,BLM}. The direct contribution of the tensor
spectrum to the $C_{\ell}$ spectrum is also of some  interest and,
similarly, we obtain the 2$\sigma$ upper limit
\begin{equation}
R_{10}< 0.32,\label{eq:R10const}
\end{equation}
and using Eq.~(\ref{eq:amp}) we obtain an upper limit on the energy
scale of inflation
\begin{equation}
\frac{H}{m_{\rm Pl}}< 1.4\times 10^{-5}.
\end{equation}
The constraints on the horizon-flow parameters are best summarized in
Fig.~\ref{fig:e1e2}. However, we can define a new parameter
along the tensor degeneracy direction and obtain the constraint
\begin{eqnarray}\label{eq:sr_constraint}
    \epsilon_2+5.0\epsilon_1 &=& 0.036 \pm0.025 \label{eq:e1e2constraint},\\
    10^{10}A_Se^{-2\tau}+82\epsilon_1 &=& 19.3\pm0.7,
\end{eqnarray}
and these constraints are displayed in Fig.~\ref{fig:e3}. To obtain
constraints on the shape of the inflaton potential we use the
slow-roll approximation
\begin{eqnarray}
\frac{H^2}{m_{\rm Pl}^2} &\simeq& \frac{8\pi}{3 m_{\rm Pl}^4} V,\\
\epsilon_1 &\simeq& \frac{m_{\rm Pl}^2}{16 \pi} \left({V'\over
V}\right)^2,\label{eq:e1}\\
\epsilon_2 &\simeq& \frac{m_{\rm Pl}^2}{4 \pi}
 \left[\left({V'\over V}\right)^2 - {V''\over V}\right],\label{eq:e22}
\end{eqnarray}
and rewriting the constraints
Eq.~(\ref{eq:eps1const})--(\ref{eq:e1e2constraint})  we find
\begin{eqnarray}
       \frac{V}{m_{\rm Pl}^4}&<& 0.23\times 10^{-10},\\
      \frac{m_{\rm Pl}^2}{16 \pi} \left({V'\over V}\right)^2&<& 0.032,\\
\frac{m_{\rm Pl}^2}{4 \pi}\left[ 2.25\left({V'\over V}\right)^2 -
{V''\over V}\right]&=& 0.036 \pm0.025\label{eq:Vconstraint}.
\end{eqnarray}
From these constraints and from Fig.~\ref{fig:e1e2}, we see that
there is no evidence of deviations from the extreme slow-roll
limit $\epsilon_i = 0$, which lies comfortably within the
$1\sigma$ contour. The best one can say is that the tensors and
tilt can be used to marginally improve the fit to the data.

In comparison with the situation before WMAP (e.g.~Ref.~\cite{infcmb}), the
principal change is a considerable tightening of the uncertainties, with the
general trend of the shrinking being towards the scale-invariant case.
This is sufficient to exclude a significant chunk of slow-roll inflation
parameter space. Of the models remaining, there is a mild preference for a red
spectral index ($n<1$) but not with any significance.
Because of the tight correlation between $\epsilon_1$
and $\epsilon_2$ there exists a class of models with blue scalar
spectra which show a mild preference for a tensor component on
large scales, as is visible in Fig~\ref{fig:nsR}. This is an area
of parameter space which has not been significantly populated with
models, as models with blue spectra tend to have negligible
tensors. These models have $\epsilon_2<0$ meaning that the fractional
kinetic energy of the inflaton is decreasing, which could
correspond to models leaving a kinetic energy dominated epoch.
These models are also somewhat `protected' by the tensor
degeneracy direction and could prove quite resistant to
observational pressure for some time to come.

We now ask how well motivated it was to stop at $\epsilon_2$, by
including one further horizon-flow parameter $\epsilon_3$. This
allows us to include a running (scale-dependence) of the scalar
spectral index, given by
\begin{equation}
\alpha_{{\rm S}} \equiv \frac{dn_{{\rm S}}}{d\ln k} = -2\epsilon_1\epsilon_2
-\epsilon_2\epsilon_3 \,.
\label{eq:alphas}
\end{equation}
and to take the expressions for the spectral indices themselves to
second-order \cite{SL2}. The effect of including this extra
parameter is shown in Fig.~\ref{fig:e3}, where it has a fairly
modest effect on the likelihood distributions for other
inflationary parameters. However $\epsilon_3$ itself is poorly
constrained, and is readily consistent with zero. We conclude that
there is clearly no motivation to include this extra parameter,
with the improved goodness-of-fit being insufficient to warrant
its inclusion.
We return to this issue in Section~\ref{sec:convcrit}.

\subsection{Constraints on power-law inflation}

If we have a specific class of inflation models in mind then we
can go beyond the constraints of Fig.~\ref{fig:e1e2}, and in this
section we examine the constraints on power-law inflation. This
remains an interesting model because the potential, once
normalized to the observed perturbation amplitude, is described by a single
parameter,
and so we can expect tighter constraints than in the case of
general slow-roll models. Power-law inflation~\cite{LM} is
expansion given by
\begin{eqnarray}
a&\propto& t^p\;,\; p>1,\\
\epsilon_1&=&{1 \over p},\;\;\epsilon_i =0,\, i\geq2\,.
\end{eqnarray}
Equivalently, in terms of conformal time we have
\begin{eqnarray}
a &\propto& |\eta|^q \;,\; -\infty <q<-1\,, \\
q&=&\frac{2}{1+3w}=-\frac{1}{1-\epsilon_1}\,.
\end{eqnarray}
We obtain the constraint on power-law inflation by reading off
the 2$\sigma$ bound on $\epsilon_1$ at the intersection with
the $\epsilon_1$ axis and find:
\begin{eqnarray}
0<&\epsilon_1&<0.019,\\
 &p &>53,\\
-1.019<&q&<-1.
\end{eqnarray}
The constraint on $\epsilon_1$ is tighter than for slow-roll
models because power-law inflation requires a red scalar power
spectrum, and so there is no possibility of taking advantage of
the tensor degeneracy. If in the near future the HZ spectrum is
ruled out in favor of red-tilted spectrum then we can begin to
place an upper limit on the index $p$, and needless to say,
power-law inflation can be ruled out altogether if a blue-tilted
spectrum is favored or if the tensor spectrum has the wrong
relationship with the scalar spectrum.

An easy way to overinterpret the data would be to point out that our
best-fit models lie close to the region well described by power-law
inflation (ie. the $\epsilon_1$ axis), with inflationary parameters
$\epsilon_1\simeq0.01$ and $\epsilon_2\simeq0.00$. However this is
likely to be a result of the mild preference for red scalar
spectra combined with a small slide along the (flattish) tensor degeneracy
direction once tensors are included in the fit. Nonetheless, it is
intriguing that such a simple inflation model should provide such an excellent
fit to the
data.

\subsection{Constraints on monomial inflation}
\label{sec:monomial}

In this section we examine the constraints on monomial inflation
potentials of the form
\begin{equation}
 V= \lambda m_{{\rm Pl}}^4 \left({\phi \over m_{{\rm
Pl}}}\right)^\alpha.\label{eq:pol_pot}
\end{equation}
Once normalized to observations these models are defined by two
parameters, the index $\alpha$ and $\phi_*$, the value of $\phi$
when the scale $k_*$ crossed the horizon during inflation.
However, there is an extra ingredient: the energy scale of
inflation is fixed once we specify $A_{{\rm S}}$ and $\epsilon_1$,
which is typically around $10^{16}$GeV. This in turn specifies the
maximum number of $e$-folds of slow-roll inflation after horizon
scale crossing, $N_{{\rm hor}}^{{\rm max}}$, after which inflation must
end (regardless of the mechanism that ends inflation) giving way
to reheating and the standard expansion history \cite{SD,efolds}.
In addition, monomial inflation provides a mechanism to end
inflation via the violation of slow-roll, and so we can calculate
the functions $H(N)$, $\epsilon_1(N)$, and $\epsilon_2(N)$ where
$N$ is the number of $e$-folds from the end of inflation. This
allows us to map a constraint on $N_{{\rm hor}}$ to a constraint
on the slow-roll parameters.

Substitution of the potential of Eq.~(\ref{eq:pol_pot}) into
Eq.~(\ref{eq:e1}) and (\ref{eq:e22}) gives
\begin{eqnarray}
\epsilon_1 &\simeq& {\alpha^2 \over 16 \pi}\left({\phi \over
m_{{\rm Pl}}} \right)^{-2},\\
\epsilon_2 &\simeq& {\alpha \over 4\pi}\left({\phi \over m_{{\rm
Pl}}} \right)^{-2} = \frac{4}{\alpha}\epsilon_1\label{eq:e2}.
\end{eqnarray}
We can calculate the number of $e$-folds of inflation from the
definition
\begin{eqnarray}\label{eq:de1}
    \frac{d\epsilon_1}{dN}&\equiv&\epsilon_1\epsilon_2,\\
    \Rightarrow N &=&
    \int^1_{\epsilon_1(N)}\frac{d\epsilon_1}{\epsilon_1\epsilon_2},\label{eq:N}
\end{eqnarray}
where ${\epsilon_1(N)}$ is the initial value of ${\epsilon_1}$.
Using $\epsilon_2=4\epsilon_1/\alpha$, which is a good approximation
for monomial inflation, we have
\begin{equation}
    N=\frac{\alpha}{4}\left[\frac{1}{\epsilon_1}-1 \right],
\end{equation}
and in the limit $N\gg\alpha/4$ the horizon flow parameters are given by
\begin{eqnarray}
\epsilon_1 &\simeq& \frac{\alpha}{4N},\\
\epsilon_2 &\simeq&\frac{1}{N}.\label{eq:e2pot}
\end{eqnarray}
Eq.~(\ref{eq:e2pot})
gives us the useful constraint
\begin{equation}
\epsilon_2> \frac{1}{N_*^{{\rm max}}}\label{eq:e2constraint}\,,
\end{equation}
where $N_*$ is the number of $e$-foldings from the end of inflation that the
scale $k_*$ crossed the horizon (in the language of Ref.~\cite{efolds}, $N_*
\simeq
N_{{\rm hor}} - 4$).
It should be emphasized that the specific form of this constraint
applies only for the monomial potentials. The maximum number of
$e$-folds of slow-roll inflation corresponding to our scale $k_*=
0.01 \, {\rm Mpc}^{-1}$ is approximately 60 \cite{SD,efolds} and
so we have the constraint $\epsilon_2> 0.017$. Using this
additional constraint we can rule out certain monomial inflation
models independent of the actual number of $e$-folds of slow-roll
inflation after horizon scale crossing.

\begin{figure}[t]
\includegraphics[width=\linewidth]{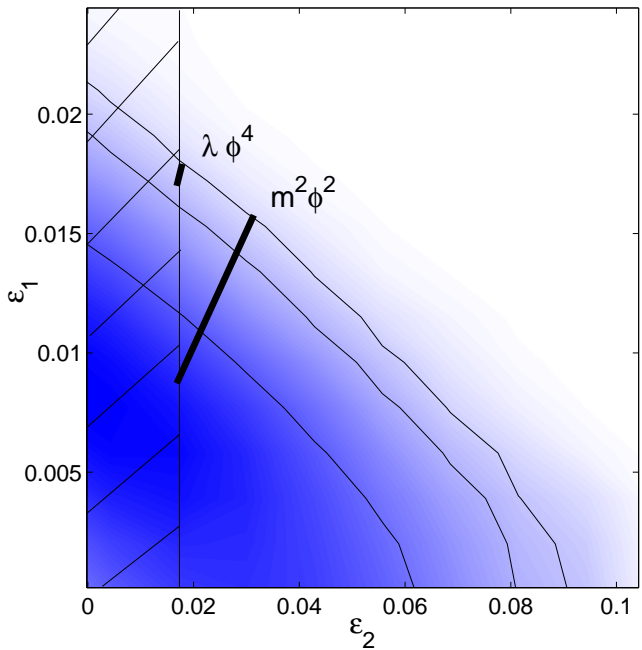}\\
\caption[fig0]{\label{fig:e1e2_zoom} 2D posterior constraints in the
$\epsilon_1$--$\epsilon_2$ plane, for the region $\epsilon_2>0$. The
contours are the $1\sigma$, $2\sigma$ and $3\sigma$ bounds. The
hatched region $\epsilon_2<1/60$ is inaccessible to monomial
inflation models. The thick lines indicate the available parameter
space for two monomial inflation models: the $\lambda \phi^4$
model is under strong pressure from observations.}
\end{figure}

In Fig.~\ref{fig:e1e2_zoom}, which is a zoom of part of
Fig.~\ref{fig:e1e2}, we restrict our attention to $\epsilon_2>0$
which corresponds to models where the ratio between the scalar
field kinetic and total energy density is increasing, arguably the
most natural candidates for ending inflation by the violation of
slow-roll. The solid lines show the location of the quadratic and
quartic potential models at different numbers of $e$-foldings up
to the maximum. The quartic potential lies outside the $2\sigma$
contour for any allowed value of $N_*$.

We can constrain the exponent $\alpha$ by reading off the value of
$4\epsilon_1\times N_*^{{\rm max}}$ at the point where the observational
constraints on the slow-roll parameters intersect with the
constraint on $\epsilon_2$ given by Eq.~(\ref{eq:e2constraint}).
We find the monomial inflation index is constrained from above to
be
\begin{eqnarray}\label{eq:alphaconst}
\alpha &<& 4.3,\;\; 3\sigma\nonumber\\
\alpha &<& 3.9,\;\; 2\sigma\nonumber\\
\alpha &<& 2.8,\;\; 1\sigma.
\end{eqnarray}
Thus, while it is still too early to definitively rule out the
$\lambda \phi^4$ inflation model, it is clear from the pattern of
the above constraints that this particular model is under very
strong pressure from observations. The $m^2 \phi^2$ is valid as
long as $32<N_{{\rm hor}}<60$ (note that monomial inflation models
with $N_{{\rm hor}}<55$ must have a prolonged reheating epoch at
the end of inflation or non-standard post-inflationary evolution \cite{efolds}).
It is clear from
Fig.~\ref{fig:e1e2_zoom} that the same kind of pressure will build
on the $m^2\phi^2$ model if the spectral index becomes more
tightly constrained, since lines of constant spectral index
correspond to $\epsilon_1= -(\epsilon_2+n_{{\rm S}}-1)/2$, which,
modulo the tensor degeneracy, runs parallel to the constraint contours.

It is natural to ask what can we say in a model-independent manner
about slow-roll inflation models while still including the bound
on $N_{{\rm hor}}$. The number of $e$-folds of inflation, given by
Eq.~(\ref{eq:N}), depends only on the function
$\epsilon_2(\epsilon_1)$, and this gives us a straightforward way
to study the phenomenology of inflation models. Unfortunately, we
only have information about the function $\epsilon_2(\epsilon_1)$
across the 8 or so $e$-folds across observable scales, and so the
rest of this function can be at best extrapolated from observable
scales. The initial slope is given by
\begin{equation}
\frac{d\epsilon_1}{d\epsilon_2}= \frac{\epsilon_1}{\epsilon_3},
\end{equation}
and so the higher slow-roll parameters, via the running of
the spectral index Eq.~(\ref{eq:alphas}), would contain useful
additional information about the model of inflation, assuming
inflation occurs at these high energies. Since we have found
no constraint on $\epsilon_3$, then it is clear that we can not
make this type of model independent analysis at present. Note also that
the bound $N_*<60$ can be
relaxed somewhat to $N_*+ \Delta N$ if $H$ is significantly reduced
in the late stages of inflation \cite{efolds}. In terms of the slow-roll
parameters
\begin{equation}
\Delta N =\ln\left(\frac{H_{{\rm i}}}{H_{{\rm f}}}\right)\simeq
\int^1_{\epsilon_1(N=60)} \frac{d\epsilon_1}{\epsilon_2}.
\end{equation}
For instance, this correction is relevant for the $\lambda\phi^4$
which has $\Delta N \simeq 4$, although this is certainly not
enough to take much pressure off this model.

\section{Discussion}
\subsection{The Harrison--Zel'dovich spectrum}

We have discounted the inclusion of $\epsilon_3$ because the data
are unable to distinguish it from zero. However, following the
logic of that statement we are forced to conclude that neither
$\epsilon_1$ nor $\epsilon_2$ have been shown to be inconsistent
with zero. The minimal assumption concerning the data is in fact
that it is due to a Harrison--Zel'dovich (HZ) spectrum, with no
indication of inflationary dynamics.

\begin{figure}[t]
\includegraphics[width=\linewidth]{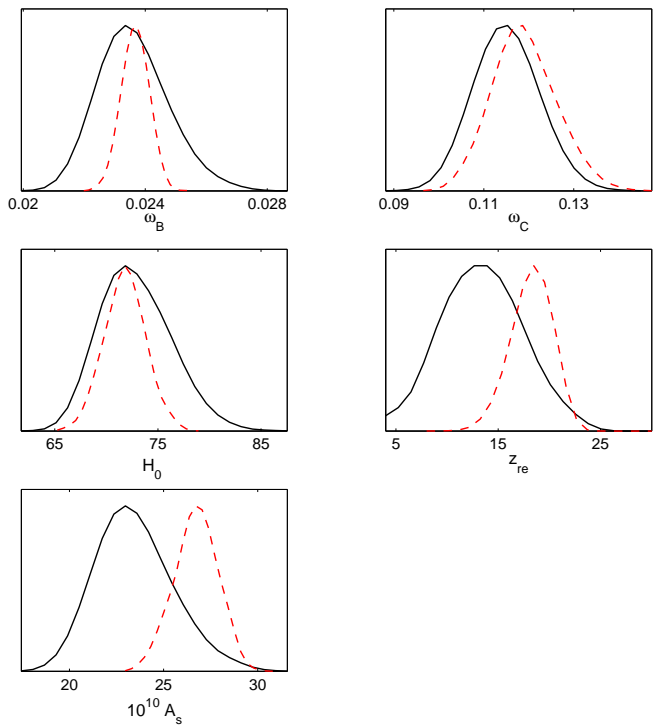}\\
\caption[fig2]{\label{fig:cos_params} 1D posterior constraints for
the basic cosmological parameters. The dotted line corresponds to
the Harrison-Zel'dovich model and the full line corresponds to
inflationary models (as shown in Fig.~\ref{fig:cos_params1}).}
\end{figure}

It is plausible therefore to also regard $\epsilon_1$ and
$\epsilon_2$ as giving poor value, in terms of the improvement to
the fit they give, for their inclusion as extra parameters. It is
therefore interesting to see how constraints tighten up if we
assume an HZ spectrum in place of slow-roll inflation. In
Fig.~\ref{fig:cos_params} we display once more the constraints on
the basic cosmological parameters, but now also showing the
constraints assuming the HZ spectrum and no tensors. It is clear
that, broadly speaking, the two models are in good agreement with
each other, but nevertheless there are quantitative differences
worth remarking upon. The baryon density $\omega_{{\rm B}}$
becomes much more tightly constrained, and looks rather high when
compared to nucleosynthesis constraints. Perhaps most
interestingly, assuming HZ the constraints on the reionization
epoch tighten considerably in favour of early reionization, as in abandoning
slow-roll we have
lost the ability to use tilt and tensors to partially mimic the
effects of reionization.

\subsection{A convergence criterion}
\label{sec:convcrit}

As long as the HZ model remains a good fit to the data then the
issue of introducing the spectral index as a parameter, let alone
the running of the spectral index, needs some consideration. Reconstructing the
initial power
spectrum in bands \cite{BLWE,MW} indicates that it
is adequately fit by a scale-invariant spectrum, and that there
should be nothing to be gained by adding any further power
spectrum parameters. Indeed, why do we include further power
spectrum parameters at all? If we consider the power spectrum as a
Taylor expansion
\begin{eqnarray}\nonumber
\label{plex} \ln {{\cal P}(k)} &=& \ln {\cal P}(k_*) + (n_{{\rm
S}}-1) \ln \left(k\over k_*\right) +\\&& \frac{\alpha_{{\rm
S}}}{2} \ln^2\left(k\over k_*\right) + \dots .\label{eq:taylor}
\end{eqnarray}
then the primary reason for including higher power spectrum
parameters is to the test the {\it convergence} of the observable,
in this case $\ln {\mathcal P}(k)$, as described at lower-order. The first term 
in the series
has been determined to be $\ln {\cal
P}(k_*)\simeq-20$. We can introduce the next order parameter, the
spectral index, without disrupting the constraints on the other
parameters, and moreover
feel confident of our measurement of $\ln {\cal P}(k_*)$, provided
\begin{equation}
\left|\ln{\mathcal P}(k_*)\right|\gg\left|(n_{{\rm S}}-1)\ln
\left({k\over k_*}\right)\right|.
\end{equation}
The above criterion is indeed satisfied by all the models under
consideration since $\max|\ln(k/k_*)|\simeq 4$ and $\max|n_{{\rm
s}}-1|\simeq 0.07$. Thus the determination of the amplitude of
scalar perturbations to a reasonable accuracy is not questioned by
anybody. Similarly, one can feel confident in a measurement of
$n_{{\rm S}}$ if the criterion
\begin{equation}
\left|n_{{\rm S}}-1\right| \gg\left|\frac{\alpha_{{\rm S}}}{2}\ln
\left(k\over k_*\right)\right|\label{eq:convcrit}
\end{equation}
is satisfied, and the fact that this inequality is typically
violated for many of the strong running models under consideration
in the literature ({\it eg.} the best fit of Ref.~\cite{wmapS} is
$n_{{\rm S}}-1=-0.07$, $\alpha_{{\rm S}}=-0.031$ and the two terms
are of the same order of magnitude) has two possible explanations.
The first is the intriguing possibility that the third term in the
Taylor expansion Eq.~(\ref{eq:taylor}) dominates over the second
term at around a scale of $\ln(k/k_*)\simeq 4$, physically
corresponding to a power spectrum with a maximum near $k_*$. This
can only be verified if the contribution to
the power spectrum from the {\it fourth} term (the running of the
running) is found to be less than the term due to the running
itself at this scale. The second, and more likely, explanation is
that the large allowed variation in the running of the spectral
index is just a symptom of the fact that we haven't convincingly
determined the spectral index yet. We suggest that this criterion
can be easily used as a check in the fitting procedure: if
Eq.~(\ref{eq:convcrit}) is violated for most models and the running is detected 
only at low significance, then the simplest
interpretation is that no useful information is coming from the
inclusion of the new parameter in the fit. One should at least be aware
of a lack of convergence in the power spectrum observable,
Eq.~(\ref{eq:taylor}).

Alternatively, one can think of Eq.~(\ref{eq:convcrit}) as setting
the boundary between weak and strong running in an observational
sense. For current observations weak running is given by models
with $|\alpha_{{\rm S}}|< \max|n_{{\rm S}}-1|/2\simeq 0.03$, and
this can be useful one wishes to introduce a weak running prior
into the fitting procedure as a perturbation analysis to the
existing constraints.

This type of convergence criterion should find echoes in other
areas of cosmology where an observable is in some way series
expanded, for instance the dark energy equation of state
$w(z)$ \cite{HT}.

Before concluding, we bring together the cosmological parameters
of various inflationary models in Table~\ref{table1}. The models
have been selected by fixing their values of $\epsilon_1$ and
$\epsilon_2$, and where a chain was not already available we have
run a short chain to determine reasonable values for the other
parameters. These models may be useful as fiducial parameters for
future study and, being within the 1$\sigma$ contour of
Fig.~\ref{fig:e1e2}, they are approximately degenerate at the
level of the current data set.

\begin{table}[t]
\begin{tabular}{l c c c c c c c| c c }
 & $\omega_{{\rm B}}$ & $\omega_{{\rm D}}$ & $H_0$ & $z_{{\rm re}}$ &
$10^{10}A_{{\rm S}}$ & $\epsilon_1$ & $\epsilon_2$ & $n_{{\rm S}}$ & $R_{10}$ \\
\hline \hline
BF  & 0.023 & 0.117 & 71.3 & 13.1 & 23.4 & 0.008 & -0.0007 & 0.98 & 0.07
\\
BFT &0.022 & 0.115 & 70.5 & 15.0 & 24.1 &0 &0.03 & 0.97 & 0
\\
BFHZ & 0.024 & 0.119 & 71.9 & 18.5 & 26.8 &0 &0 & 1 & 0
\\
$\lambda\phi^4$ &0.022 &0.107 &71.5 & 7.1 & 20.3 & 0.017 & 0.017 &
0.95 & 0.13
\\
$m^2\phi^2$&0.023 &0.114 & 70.9& 10.5 & 22.1 & 0.008 & 0.017 & 0.97 & 0.06
\\
TD1$\sigma$ &0.025 &0.107 & 77.4 &15.0 &23.0 &0.023 & -0.077 & 1.03 & 0.21
\\
B1$\sigma$ & 0.025 & 0.113 &76.7 & 17.2 & 24.7 & 0.018 & -0.077 & 1.04
& 0.16
\\
\hline \hline
\end{tabular}
\caption{\label{table1}Cosmological parameters for various
inflationary models selected by their values of $\epsilon_1$ and
$\epsilon_2$, with power spectrum parameters defined at $k_*
=0.01\mbox{Mpc}^{-1}$. The models are the overall best-fit (BF),
the best-fit tilted (BFT), Harrison-Zel'dovich (BFHZ), the
$\lambda\phi^4$ and $m^2\phi^2$ 60 $e$-fold models, the best-fit
model at the tip of the 1$\sigma$ contour along the tensor
degeneracy (TD1$\sigma$), and the best-fit model with the bluest
scalar spectrum along the 1$\sigma$ contour (B1$\sigma$).}
\end{table}

\section{Conclusions}

We have investigated the constraints on various slow-roll
inflationary models coming from observations of the CMB and
large-scale structure. The main result is that the viable
slow-roll parameter space is dramatically reduced and the
underlying inflationary degeneracy now becomes visible.
Interestingly, if we combine these constraints with a constraint
on the number of $e$-folds of inflation since horizon scale
crossing, then we find that the $\lambda \phi^4$ inflation model
is under strong pressure, though not yet definitively ruled out.
The $m^2\phi^2$ model will come under the same threat as long as
the data continue to favour the Harrison-Zel'dovich spectrum.

We also introduced a simple convergence criterion,
Eq.~(\ref{eq:convcrit}), to judge the necessity of including
higher power spectrum parameters such as the spectral index and
the running of the spectral index. Applying this criterion we find
that while it is justified to include the spectral index in the
fit (reflecting the fact that amplitude of scalar perturbations is
now well determined), it is not useful to include the running of
the spectral index at present.

Our inflation analysis comes after those of Refs.~\cite{wmapinf,BLM,KKMR}, with
whom we find general agreement. We can make the most direct comparison
with Barger {\it et al.}~\cite{BLM}, who used the WMAP data alone with
a top-hat $H_0$ prior, and a grid based $\chi^2$ maximization procedure
to obtain their constraints. We have have found tighter constraints
using WMAP, VSA, CBI, ACBAR, 2dF datasets, and an MCMC technique, with
our 2$\sigma$ constraint in Fig.~\ref{fig:e1e2} being only slightly
looser than their equivalent 1$\sigma$ constraint. The comparison with
Peiris {\it et al.}~\cite{wmapinf} is less straightforward, since we did
not consider the effect of strong running of the spectral index,
arguing in Sec.~\ref{sec:convcrit} that we can obtain better value from
the current data set without it. As a result, their results marginally
favour a model with a blue scalar spectrum on the largest scales (even in the
tensorless
limit), whereas the power-law fits in the literature, including their
adiabatic/isocurvature fit, favour a slightly red spectrum. Both Peiris
{\it et al.} and Kinney {\it et al.}~\cite{KKMR} investigate the
Monte-Carlo flow reconstruction technique, which inevitably involves
an extrapolation of the inflationary potential and physics well beyond
the region directly constrained by observations. Therefore a certain
caution is required when interpreting those analyses. We have hinted
in Sec.~\ref{sec:monomial} how one might make such an extrapolation of
the horizon-flow parameters (instead of the potential), but the method
would require considerably tighter constraints on the running of the
spectral index than we have at present.


\begin{acknowledgments}
S.M.L is supported by the EU CMBNET network, and A.R.L.~in part by
the Leverhulme Trust. We thank Dominik Schwarz, Ruth Durrer,
Roberto Trotta, and Micha\"el Malquarti for useful discussions.
\end{acknowledgments}



\begin{thebibliography}{}

\bibitem{wmap1} C. L. Bennett {\it et al.}, astro-ph/0302207.
\bibitem{wmapS} D. N. Spergel {\it et al.}, {\tt astro-ph/0302209}.
\bibitem{wmapinf} H. V. Peiris {\it et al.}, astro-ph/0302225.
\bibitem{wmap2} E. Komatsu {\it et al.}, astro-ph/0302223.
\bibitem{wmap3} G. Hinshaw {\it et al.}, {\tt astro-ph/0302217}.
\bibitem{wmap4} A. Kogut {\it et al.}, astro-ph/0302213.
\bibitem{LLMS} S. M. Leach, A. R. Liddle, J. Martin, and  D. Schwarz,
    Phys.
Rev. D{\bf 66} 023515 (2002), {\tt astro-ph/0202094}.
\bibitem{infcmb} S. M. Leach and A. R. Liddle, Mon. Not. Roy. Astr.
    Soc. {\bf 341} 1151, (2003), {\tt astro-ph/0207213}.
\bibitem{MCMC} N. Christensen and R. Meyer, {\tt astro-ph/0006401};
    N. Christensen, R. Meyer, L. Knox, and B. Luey, Class.
    Quant. Grav. {\bf 18}, 2677 (2001), {\tt astro-ph/0103134}.
\bibitem{LB} A. Lewis and S. Bridle, Phys. Rev. D{\bf 66}, 103511 (2002),
    {\tt astro-ph/0205436}.
\bibitem{Sel} U. Seljak, P. MacDonald, and A. Makarov, {\tt astro-ph/0302571}.
\bibitem{BLM} V. Barger, H. Lee, and D. Marfatia, {\tt hep-ph/0302150}.
\bibitem{KKMR} W. H. Kinney, E. W. Kolb, A. Melchiorri, and A. Riotto,
    {\tt hep-ph/0305130}.
\bibitem{SL} V. F. Mukhanov and G.~V.~Chibisov, JETP Letters, {\bf
    33}, 532 (1981), {\tt astro-ph/0303077};
V. F. Mukhanov, H. A.
    Feldman, and R. H. Brandenberger, Phys. Rep. {\bf 215}, 203
    (1992).
\bibitem{SL2} E. D. Stewart and D.~H.~Lyth, Phys. Lett. B, {\bf 302},
    171 (1993), {\tt gr-qc/9302019}.
\bibitem{vsa} K. Grainge, {\it et al.}, {\tt astro-ph/0212495}.
\bibitem{cbi} T. J. Pearson {\it et al.}, {\tt astro-ph/0205388}.
\bibitem{acbar} C. L. Kuo {\it et al.}, {\tt astro-ph/0212289}.
\bibitem{2df} W. J. Percival {\it et al.}, Mon. Not. Roy. Astr. Soc.
    {\bf 327}, 1297 (2001), {\tt astro-ph/0105252}.
\bibitem{LCL} A. Lewis, A. Challinor, and A. Lasenby, Astrophys. J. {\bf 538},
    473 (2000), {\tt astro-ph/9911177}.
\bibitem{V} L. Verde {\it et al.}, astro-ph/0302218.
\bibitem{LM} F. Lucchin and S. Matarrese, Phys. Rev. D {\bf 32}, 1316 (1985).
\bibitem{SD} S. Dodelson and L. Hui, {\tt astro-ph/0305113}.
\bibitem{efolds} A. R. Liddle and S. M. Leach, {\tt astro-ph/0305263}.
\bibitem{BLWE}  S. L. Bridle, A. M. Lewis, J. Weller, and
G. Efstathiou,
    {\tt astro-ph/0302306}.
\bibitem{MW} P. Mukherjee and Y. Wang, {\tt astro-ph/0303211}.
\bibitem{HT} D. Huterer and M. S. Turner, AIP Conf. Proc. {\bf 599}, 140 (2001) 
	{\tt
astro/ph 0006419};
I. Maor, R. Brustein, and P. J. Steinhardt,
	Phys. Rev. Lett. {\bf 86}, 6 (2001), Erratum-ibid. {\bf 87}, 049901
	(2001) {\tt astro-ph/0007297};
J. Weller and A. Albrecht, Phys. Rev.
	D{\bf 65},103512
(2002) 103512, {\tt astro-ph/0106079}
7222 (1994), {\tt astro-ph/9408015}.

\end{thebibliography}
\end{document}